\documentclass[twocolumn]{aastex631}

\usepackage{color}

\shorttitle{Light Curves and Event Rates of AISNe}
\shortauthors{Mori et al.}
\graphicspath{{./}{figures/}}

\begin{document}

\title{Light Curves and Event Rates of Axion Instability Supernovae}

\author[0000-0003-2595-1657]{Kanji Mori}
\affiliation{Research Institute of Stellar Explosive Phenomena, Fukuoka University,\\ 8-19-1 Nanakuma, Jonan-ku, Fukuoka-shi, Fukuoka 814-0180, Japan}
\email{kanji.mori@fukuoka-u.ac.jp}

\author[0000-0003-1169-1954]{Takashi J. Moriya}
\affiliation{National Astronomical Observatory of Japan, 2-21-1 Osawa, Mitaka, Tokyo 181-8588, Japan}
\affiliation{School of Physics and Astronomy, Faculty of Science, Monash University, Clayton, Victoria 3800, Australia}

\author[0000-0003-0304-9283]{Tomoya Takiwaki}
\affiliation{National Astronomical Observatory of Japan, 2-21-1 Osawa, Mitaka, Tokyo 181-8588, Japan}
\author[0000-0003-2456-6183]{Kei Kotake}
\affiliation{Research Institute of Stellar Explosive Phenomena, Fukuoka University,\\ 8-19-1 Nanakuma, Jonan-ku, Fukuoka-shi, Fukuoka 814-0180, Japan}
\affiliation{Department of Applied Physics, Faculty of Science, Fukuoka University,\\ 8-19-1 Nanakuma, Jonan-ku, Fukuoka-shi, Fukuoka 814-0180, Japan}
\author[0000-0001-6142-6556]{Shunsaku Horiuchi}
\affiliation{Center for Neutrino Physics, Department of Physics, Virginia Tech, Blacksburg, VA 24061, USA}
\affiliation{Kavli IPMU (WPI), UTIAS, The University of Tokyo, Kashiwa, Chiba 277-8583, Japan}
\author[0000-0002-5726-538X]{Sergei I. Blinnikov}
\affiliation{National Research Center ``Kurchatov Institute", 123182 Moscow, Russia}
\affiliation{Kavli IPMU (WPI), UTIAS, The University of Tokyo, Kashiwa, Chiba 277-8583, Japan}



\begin{abstract}

It was recently proposed that exotic particles can trigger a new stellar instability which is analogous to the $e^-e^+$ pair instability if they are produced and reach equilibrium in the stellar plasma. In this study, we construct axion instability supernova (AISN) models caused by the new instability to predict their observational signatures. We focus on heavy axion-like particles (ALPs) with masses of $\sim 400$\,keV--$2$\,MeV and coupling with photons of $g_{a\gamma}\sim10^{-5}$\,GeV$^{-1}$. It is found that the $^{56}$Ni mass and the explosion energy are significantly increased by ALPs for a fixed stellar mass. As a result, the peak times of the light curves of AISNe occur earlier than those of standard pair-instability supernovae by 10--20 days when the ALP mass is equal to the electron mass. Also, the event rate of AISNe is $1.7$--$2.6$ times higher than that of pair-instability supernovae, depending on the high mass cutoff of the initial mass function.

\end{abstract}



\section{Introduction} \label{sec:intro}
The equation of state (EoS) of plasma determines stellar structure and evolution.  In the hot plasma with temperatures $T\gtrsim 10^9$\,K and the densities $\rho<10^6$\,g\,cm$^{-3}$, the $e^-e^+$ pair-creation takes place and the EoS becomes softer if photons are in thermodynamic equilibrium with the plasma. As a result, in the carbon-oxygen cores of very massive stars with initial mass of $M_\mathrm{init}\sim150$--$260M_\odot$, the adiabatic index $\Gamma_1$ becomes lower than 4/3 and the star becomes dynamically unstable. The core then contracts and explosive oxygen burning releases a large amount of energy to disrupt the whole star \citep{1967SvA....10..604B,1967ApJ...148..803R,1967PhRvL..18..379B,1968Ap&SS...2...96F}. The stellar explosion caused by the $e^-e^+$ pair-creation is called a pair instability supernova (PISN).

The extremely hot environment in PISNe could be used as a laboratory  to probe fundamental physical processes such as new particles \citep{2020PhRvD.102k5024C,2020PhRvL.125z1105S,2022PhRvD.105i5038S}, the $^{12}$C$(\alpha, \gamma)^{16}$O reaction rate \citep{2018ApJ...863..153T,2020ApJ...902L..36F}, and relativistic plasma screening \citep{2022A&A...659A..97F}. These studies were motivated by the recent gravitational-wave detection of black hole mergers \citep[e.g.,][]{2021arXiv211103634T}. When the stellar mass is $M_\mathrm{init}\sim80$--$150M_\odot$, the star is not totally disrupted but a large part of its mass can be ejected by the pulsational pair instability. As a result, the so-called pair-instability mass gap of black holes is formed \citep[e.g.,][]{2016A&A...588A..50M,2017ApJ...836..244W,2019ApJ...887...72L,2019ApJ...887...53F,2021ApJ...910...30T,2021MNRAS.501L..49K}. Because the position of  the lower edge of the mass gap could be estimated from a large sample of black hole mergers, it is possible to compare the observations and stellar models to obtain information on fundamental physics. 

Recently, \citet{2022PhRvD.105i5038S} developed stellar models which consider new bosons beyond the Standard Model. If the new boson is as massive as or lighter than $\sim2m_e$, where $m_e$ is the electron mass, and is strongly coupled with the plasma, it can be produced in PISNe and reach equilibrium. In such a case, the EoS is softened and a new stellar instability can be induced. \citet{2022PhRvD.105i5038S} focused on the pair-instability mass gap and found that the lower edge of the mass gap decreases by $\sim10M_\odot$ when the mass of the new particle is equal to $m_e$. Although their method is applicable to any bosons, they focused on heavy axion-like particles (ALPs) that interact with photons. They coined the term ``axion instability supernova" (AISN) for the new transient that is induced by ALPs.

PISNe and pulsational PISNe have been linked to observed superluminous supernovae such as SN 2007bi \citep{2009Natur.458..865G}, OGLE14-073 \citep{2018MNRAS.479.3106K}, and SN 2006gy \citep{2007Natur.450..390W}.  Also, PISNe are a target of next-generation near-infrared telescopes such as the Nancy Grace Roman Space Telescope \citep{2022ApJ...925..211M}, Euclid space telescope \citep{2022arXiv220408727M,2022arXiv220409402T}, and James Webb Space Telescope \citep{2018MNRAS.479.2202H,2020ApJ...894...94R}. These observation may discover PISN-like events and enable one to probe their nature. It is hence important to predict observable quantities beforehand.

Heavy ALPs which couple with photons have been constrained by beam dump experiments \cite[e.g.,][]{2017JHEP...12..094D} and astrophysical phenomena including horizontal branch stars \citep{2020PhLB..80935709C,2022arXiv220301336L} and core-collapse supernovae \citep{2018PhRvD..98e5032J,2020JCAP...12..008L,2022PhRvD.105f3009M,2022PhRvL.128v1103C,2022PhRvD.105c5022C}. There is an interesting region at $m_a\sim400$\,keV--$2$\,MeV and $g_{a\gamma}\sim10^{-5}$\,GeV$^{-1}$ in the ALP parameter space, where $m_a$ is the ALP mass and $g_{a\gamma}$ is the coupling constant between ALPs and photons. This region is called the ``cosmological triangle'' \citep{2020PhRvL.124u1804D} because  cosmological phenomena have been often utilized to exclude it. Recently, it was pointed out that energetics of core-collapse supernovae are likely to provide a constraint for the cosmological triangle \citep{2022PhRvL.128v1103C,2022PhRvD.105c5022C}. Nevertheless, it is desirable to access this region with independent astrophysical argument to evade  systematic uncertainties. In the cosmological triangle, the ALP-photon coupling is so strong that ALPs are confined in the stellar core. Since ALPs in this region reach equilibrium with the stellar plasma, they can affect the EoS and lead to AISNe \citep{2022PhRvD.105i5038S}. In this paper, we explore observable the signatures of AISNe caused by ALPs in the cosmological triangle.

This paper  is organized as follows. In Section 2, we describe the setup of our PISN and AISN models. In Section 3, we show the $^{56}$Ni mass, the explosion energy, the light curves, and the event rates inferred from the models. In Section 4, the future detection prospect is discussed and the paper is concluded.

\section{Method}
We use Modules for Experiments in Stellar Astrophysics \citep[MESA;][]{Paxton2011,Paxton2013,Paxton2015,Paxton2018,Paxton2019} version 12778 to calculate our one-dimensional stellar models. Input parameters such as the mixing length and overshooting follow the prescription in \cite{2022PhRvD.105i5038S}. Although it solves hydrostatic evolution for almost all stages of the stellar life, MESA is capable of switching to the HLLC solver to simulate hydrodynamical evolution of shocks and pulsations \citep{Paxton2018}. In our models, the hydrodynamical evolution is solved during the (pulsational) pair instability.

We start our simulations from  helium stars with different masses between $46M_\odot$ and $139M_\odot$ and the metallicity of $Z=10^{-5}$. We then follow their evolution until AISNe or core collapse and calculate observable quantities including the $^{56}$Ni mass and event rates. Also, we calculate models that start from zero-age main sequence (ZAMS) stars with hydrogen envelopes and follow their evolution until central hydrogen depletion to obtain the relation between the ZAMS mass $M_\mathrm{init}$ and the helium core mass $M_\mathrm{He}$. This relation enables us to calculate the event rates.

When new bosons are produced and trapped in the stellar plasma, the EoS is modified. The pressure $P_a$, density $\rho_a$, the internal energy $u_a$, and the specific entropy $s_a$ induced by ALPs are given as 
\begin{eqnarray}
    P_a&=&\frac{1}{2}m_ac^2 CH_1,\\
    \rho_a&=&\frac{1}{2}m_aCH_2,\\
    u_a&=&\frac{1}{2}m_ac^2H_3,\\
    s_a&=&\frac{kC\beta}{2\rho}(H_1+H_3),
\end{eqnarray}
where $k$ is the Boltzmann constant, $c$ is the speed of light, $C=(m_ac/\hbar)^3/\pi^2$, $\hbar$ is the Planck constant, $\beta=m_ac^2/kT$, 

\begin{eqnarray}
    H_1&=&\int^\infty_\beta G\left(\frac{\epsilon}{\beta}\right)B(\epsilon)\frac{d\epsilon}{\beta},\\
    H_2&=&\int^\infty_\beta G'\left(\frac{\epsilon}{\beta}\right)B(\epsilon)\frac{d\epsilon}{\beta},\\
    H_3&=&\int^\infty_\beta \epsilon G'\left(\frac{\epsilon}{\beta}\right)B(\epsilon)\frac{d\epsilon}{\beta^2},
\end{eqnarray}
$B(\epsilon)=(e^\epsilon-1)^{-1}$, and $G(x)=(x^2-1)^\frac{3}{2}/3$. The modification to other thermodynamic quantities stems from these terms as tabulated in \cite{2022PhRvD.105i5038S}. 

In this study, we focus on photophilic ALPs in the cosmological triangle. We adopt two ALP masses of $f=m_a/2m_e=0.5$ and 2. Here, the ALP mass is normalized to $2m_e$ because $m_a\lesssim 2m_e$ is the critical condition for the new stellar instability. Since the cosmological triangle is located at  $m_a\sim400$\,keV--$2$\,MeV \citep{2022arXiv220301336L}, the two cases approximately correspond to the lower and the upper limits of $m_a$. Also, we calculate standard PISN models for comparison.
\section{Results}
\begin{table*}[h]
\caption{The stellar models developed in this study. In the table, $f=m_a/2m_e$ is the ALP mass, $M_\mathrm{init}$ is the intial stellar mass, $M_\mathrm{He}$ is the helium core mass, $M_\mathrm{BH}$ is the black hole mass, and $M_\mathrm{Ni}$ is the $^{56}$Ni mass in the ejecta.  The first column is filled with ``$-$" for the standard models without ALPs. ``PAISN" and ``PPISN" in the seventh column stand for the pulsational AISN and the pulsational PISN, respectively. $M_\mathrm{Ni}$ and $E_\mathrm{exp}$ for PPISNe and PAISNe are not shown because they can cause multiple pulses. \label{tab:tab1}}
\centering
\begin{tabular}{cc|ccccc}
$f$ & $M_\mathrm{init}/M_\odot$ & $M_\mathrm{He}/M_\odot$ & $M_\mathrm{BH}/M_\odot$ & $M_\mathrm{Ni}/M_\odot$&$E_\mathrm{exp}$ [$10^{51}$ erg]&Fate \\\hline\hline
0.5         & 100                       & 46                    & 26.2                  &     &&PAISN                  \\
0.5         & 110                       & 51                    & 0                     & 1.07$\times10^{-3}$&4.6&AISN              \\
0.5         & 120                       & 56                    & 0                     & 2.01$\times10^{-2}$  &9.0&AISN             \\
0.5         & 130                       & 59                    & 0                     & 8.24$\times10^{-2}$ &13&AISN              \\
0.5         & 140                       & 67                    & 0                     & 4.81$\times10^{-1}$ &23&AISN              \\
0.5         & 150                       & 72                    & 0                     & 1.58      &29&AISN         \\
0.5         & 160                       & 77                    & 0                     & 4.16       &35&AISN      \\
0.5         & 170                       & 83                    & 0                     & 9.36     &43&AISN         \\
0.5         & 180                       & 88                    & 0                     & 15.0     &49&AISN          \\
0.5         & 190                       & 94                    & 0                     & 23.8     &57&AISN          \\
0.5         & 200                       & 98                    & 0                     & 31.3     &63&AISN          \\
0.5         & 210                       & 102                   & 0                     & 40.9     &69&AISN          \\
\hline
2.0         & 140                       & 67                    & 15.3              &              &&PAISN         \\
2.0         & 150                       & 72                    & 0                     & 2.20$\times10^{-2}$ &7.6&AISN              \\
2.0         & 160                       & 77                    & 0                     & 7.95$\times10^{-2}$  &12&AISN             \\
2.0         & 170                       & 83                    & 0                     & 3.12$\times10^{-1}$      &20&AISN         \\
2.0         & 180                       & 88                    & 0                     & 8.78$\times10^{-1}$  &28&AISN             \\
2.0         & 190                       & 94                    & 0                     & 2.80     &36&AISN          \\
2.0         & 200                       & 98                    & 0                     & 5.21      &42&AISN         \\
2.0         & 210                       & 102                   & 0                     & 8.26     &47&AISN          \\
2.0         & 220                       & 109                   & 0                     & 15.9     &58&AISN          \\
2.0         & 230                       & 115                   & 0                     & 26.2     &68&AISN          \\
2.0         & 240                       & 122                   & 0                     & 43.3    &82&AISN           \\
\hline
$-$       & 140                       & 67                    & 30.4              &       &&  PPISN              \\
$-$       & 150                       & 72                    & 0                     & 1.71$\times10^{-2}$   & 7.1&PISN           \\
$-$       & 160                       & 77                    & 0                     & 6.73$\times10^{-2}$      &11&PISN        \\
$-$       & 170                       & 83                    & 0                     & 2.42$\times10^{-1}$  &19&PISN            \\
$-$       & 180                       & 88                    & 0                     & 6.74$\times10^{-1}$   &27&PISN           \\
$-$       & 190                       & 94                    & 0                     & 2.09         &35&PISN     \\
$-$       & 200                       & 98                    & 0                     & 3.97         &40&PISN     \\
$-$       & 210                       & 102                   & 0                     & 6.44        &48&PISN      \\
$-$       & 220                       & 109                   & 0                     & 12.6        &54&PISN      \\
$-$       & 230                       & 115                   & 0                     & 20.1        &62&PISN      \\
$-$       & 240                       & 122                   & 0                     & 32.6         &73&PISN     \\
$-$       & 250                       & 127                   & 0                     & 43.5       &83&PISN       \\
\end{tabular}
\end{table*}
\subsection{Model Properties}
The stellar  models developed in this study are listed in Table~\ref{tab:tab1}. When the stellar mass is not sufficiently high, the star experiences  the pulsational pair instability or pulsational axion instability.  In these models, the instability causes stellar pulsations which eject a part of the stellar mass. As a result, a black hole lighter than $M_\mathrm{He}$ is left after the core collapse. \cite{2022PhRvD.105i5038S} reported that the dynamics of the pulsation is significantly affected by ALPs and the lower edge of the black hole mass gap becomes lighter. They also reported that the boundary between the pulsational instabilities and the total disruption of the star becomes lighter. In our models, the boundary is at $M_\mathrm{init}\approx145M_\odot$ for the standard case and at $M_\mathrm{init}\approx105M_\odot$ for the $f=0.5$ case.

On the other hand, when the star is heavy enough, the energy release of oxygen burning is not sufficient to stop the stellar contraction. In this case, the total energy of the star is absorbed by photodisintegration. As a result, the star directly collapses to a black hole \cite[e.g.,][]{2016MNRAS.456.1320T}. This defines the upper edge of the pair-instability mass gap of black holes. \cite{2022PhRvD.105i5038S} showed that the upper edge becomes lighter if ALPs are included. In our calculation, the boundary between the total disruption and the direct collapse is lowered from $M_\mathrm{init}\approx265M_\odot$ to $M_\mathrm{init}\approx215M_\odot$ when $f=0.5$ is assumed. 

\begin{figure}[t]
 \centering
 \includegraphics[width=8.5cm]
      {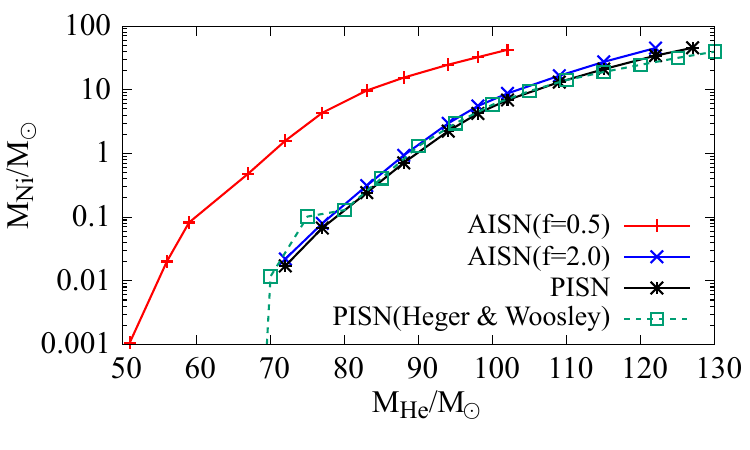}
 \caption{The $^{56}$Ni mass $M_\mathrm{Ni}$ ejected from AISNe and PISNe as a function of the helium core mass $M_\mathrm{He}$. The solid lines show our models and the dotted line shows PISN models in \cite{2002ApJ...567..532H}.
 }
 \label{Ni}
\end{figure}

\begin{figure}[t]
 \centering
 \includegraphics[width=8.5cm]
      {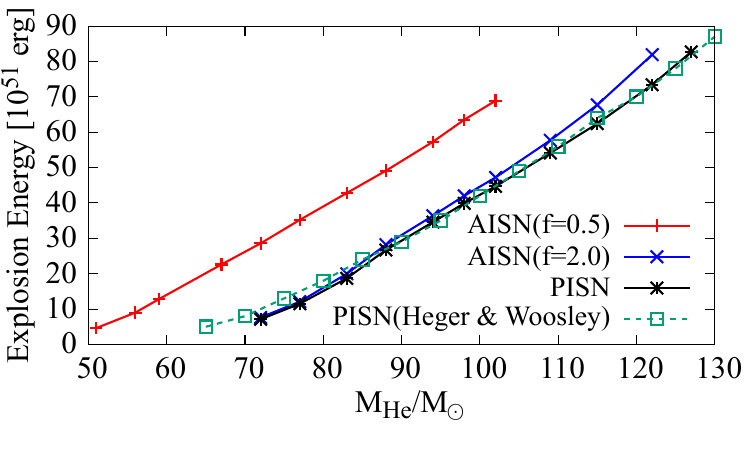}
 \caption{The explosion energies $E_\mathrm{exp}$ for AISN and PISN models as a function of the helium core mass $M_\mathrm{He}$. The solid lines show our models and the dotted line shows PISN models in \cite{2002ApJ...567..532H}.}
 \label{Eexp}
\end{figure}

Figure \ref{Ni} shows the ejected $^{56}$Ni mass $M_\mathrm{Ni}$ of AISNe and PISNe, which is defined as
\begin{eqnarray}
    M_\mathrm{Ni}=\int X_{^{56}\mathrm{Ni}} dM_r,
\end{eqnarray}
where $X_{^{56}\mathrm{Ni}}$ is the mass fraction of $^{56}$Ni and $M_r$ is the mass coordinate.  $M_\mathrm{Ni}$ increases as a function of the stellar mass for a fixed $m_a$ because the temperature in the stellar core is higher in heavier stars. This is consistent with previous PISN models \citep{2002ApJ...567..532H,2011ApJ...734..102K,2017ApJ...846..100G,2016MNRAS.456.1320T,2018ApJ...857..111T}. On the other hand, when the stellar mass is fixed, $M_\mathrm{Ni}$ in AISNe is significantly larger than that in PISNe. This is because ALPs soften the EoS and  the stellar contraction lasts until the central temperature becomes higher than standard PISNe. The iron group elements are produced in the innermost region where the temperature exceeds $\sim4.5\times10^9$\,K \citep{2018ApJ...857..111T}.

Figure \ref{Eexp} shows the explosion energies $E_\mathrm{exp}$ for the AISN and PISN models. As reported in the previous studies \cite[e.g.,][]{2002ApJ...567..532H,2016MNRAS.456.1320T},  $E_\mathrm{exp}$ increases  as a linear function of $M_\mathrm{He}$. PISNe and AISNe are more energetic than typical type II supernovae. Even the lightest PISN model with $M_\mathrm{He}=72M_\odot$ reaches $E_\mathrm{exp}=7.1\times10^{51}$\,erg. Such energetic explosions are induced by explosive oxygen burning, which is ignited when the central temperature reaches $\sim3\times10^9$\,K. Since the softened EoS induces higher temperature in the core, $E_\mathrm{exp}$ in AISNe are higher than the corresponding values in PISNe for a fixed $M_\mathrm{He}$.

\subsection{Light Curves}
\begin{figure}[t]
 \centering
 \includegraphics[width=8.5cm]
      {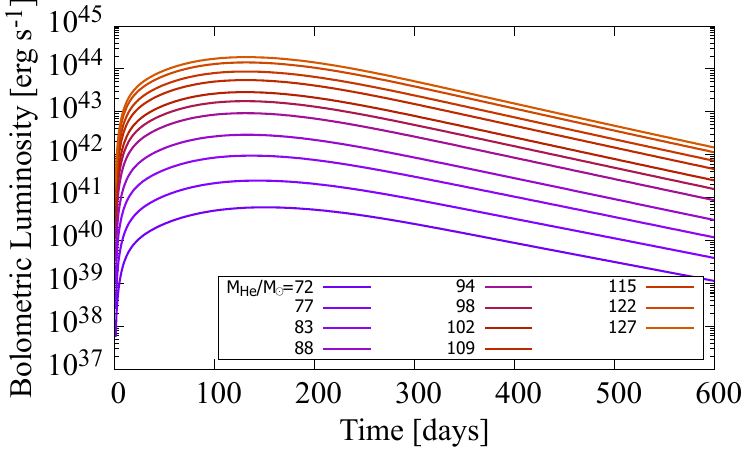}
       \includegraphics[width=8.5cm]
      {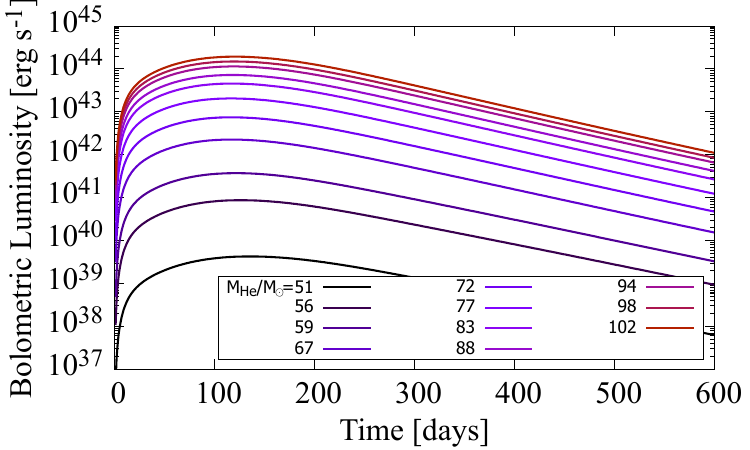}
 \caption{The bolometric light curve of PISNe (upper panel) and AISNe (lower panel) with $f=0.5$.}
 \label{LC}
\end{figure}
Although confirmed light curves of PISNe have not been observed, they may be discovered by future observations and provide information on the ejecta mass $M_\mathrm{ej}$, $M_\mathrm{Ni}$, and $E_\mathrm{exp}$ through the comparison with the Arnett law \citep{1982ApJ...253..785A} or hydrodynamical simulations \cite[e.g.,][]{2011ApJ...734..102K}. If the progenitor of PISNe and AISNe is hydrogen-free, the stellar mass $M_\mathrm{He}$ can be estimated from $M_\mathrm{ej}$. 

Figure \ref{LC} shows the light curves of PISNe and AISNe with $f=0.5$ estimated as \citep{1982ApJ...253..785A}
\begin{eqnarray}
    L(t)=M_\mathrm{Ni}e^{-x^2}\left((s_\mathrm{Ni}-s_\mathrm{Co})\int^x_02ze^{z^2-2zy}dz+\right.\nonumber\\
    \left.s_\mathrm{Co}\int 2ze^{z^2-2yz+2zs}dz\right)
    (1-e^{-A_\gamma t^{-2}}),
\end{eqnarray}
where $s_\mathrm{Ni}$ and $s_\mathrm{Co}$ are the energy generation rates of the $^{56}$Ni and $^{56}$Co decays, $x=t/t_\mathrm{eff}$, $y=t_\mathrm{eff}/2\tau_\mathrm{Ni}$, $s=t_\mathrm{eff}(\tau_\mathrm{Co}-\tau_\mathrm{Ni})/2\tau_\mathrm{Ni}\tau_\mathrm{Co}$, and $A_\gamma=(3\kappa_\gamma M_\mathrm{ej})/(4\pi v^2)$. Here $t_\mathrm{eff}=\sqrt{2t_\mathrm{d}t_\mathrm{h}}$ is the effective diffusion timescale derived from the diffusion timescale $t_\mathrm{d}$ and the dynamical timescale $t_\mathrm{h}$, $\tau_\mathrm{Ni}$ and $\tau_\mathrm{Co}$ are the $^{56}$Ni and $^{56}$Co lifetimes, $\kappa_\gamma$ is the $\gamma$-ray opacity, and $v$ is the photospheric velocity. 
\begin{figure}[t]
 \centering
 \includegraphics[width=8.5cm]
      {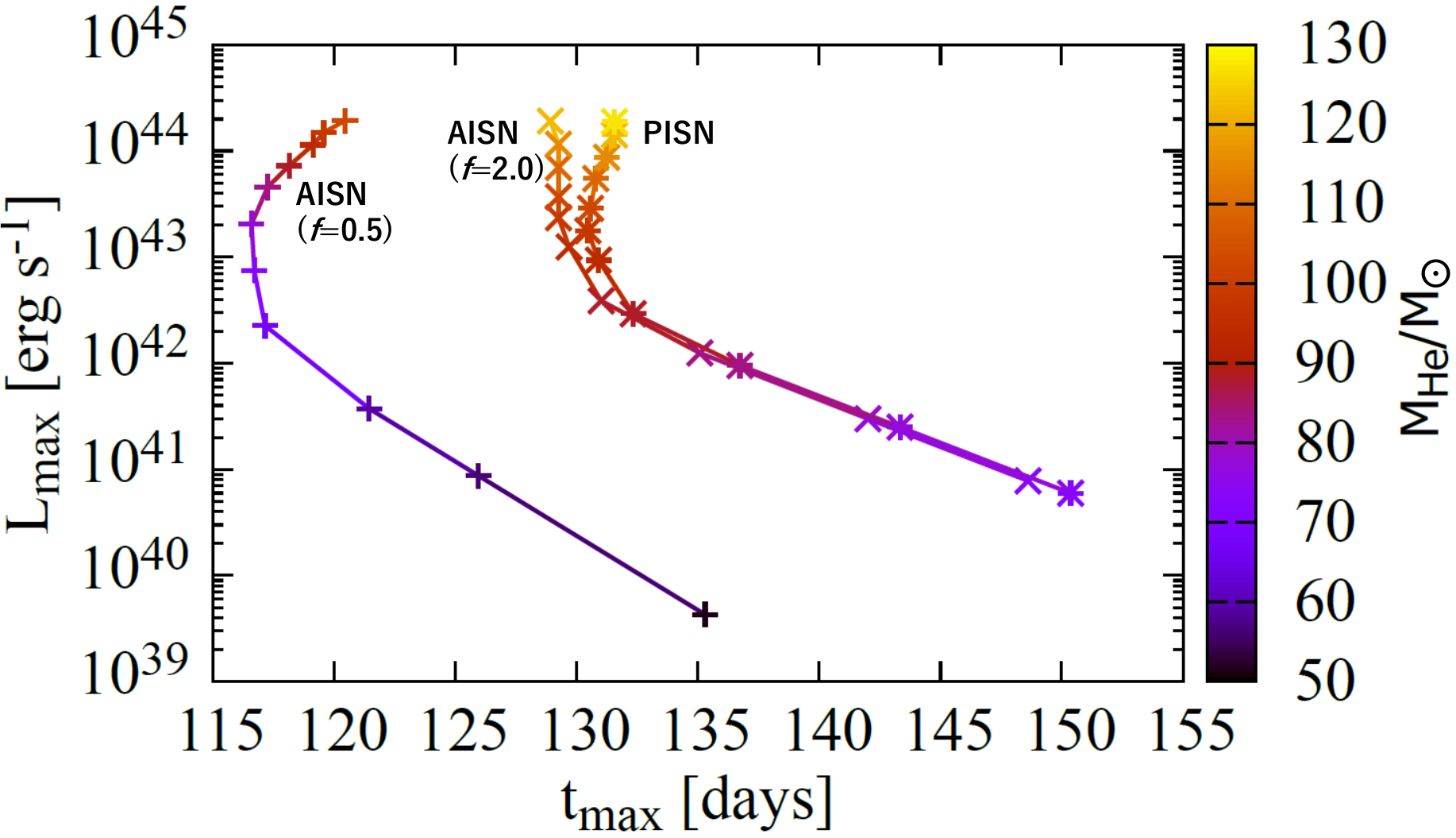}
 \caption{The peak luminosity $L_\mathrm{max}$ and the peak time $t_\mathrm{max}$ for the PISN and AISN models. The color of each point shows the helium core mass $M_\mathrm{He}$.}
 \label{LCpeak}
\end{figure}

From Fig.~\ref{LC}, it is seen that the peak luminosity $L_\mathrm{max}$ increases as a function of $M_\mathrm{He}$. This is because the luminosity is powered by $^{56}$Ni, which is produced more abundantly in heavier models. Although the heaviest models can be as luminous as $\sim10^{44}$\,erg\,s$^{-1}$, some of lighter models are less luminous than the typical supernova luminosity $\sim10^{42}$\,erg\,s$^{-1}$. The peak time $t_\mathrm{max}$ at which the luminosity reaches $L_\mathrm{max}$ is 100--150 days. The luminosity is low in the early days of $t<t_\mathrm{max}$ because the ejecta is optically thick, while it decreases  in the later days because the heating rate becomes lower. The timescale $t_\mathrm{max}$ is determined by the condition $t_\mathrm{d}\approx t_\mathrm{h}$ \citep[e.g., Chapter 5 in][]{2017suex.book.....B}. It is easily shown that this condition leads to the relation $t_\mathrm{max}\propto(M_\mathrm{ej}^3/E_\mathrm{exp})^4$. 

Since the light curves depend on the supernova properties, they provide information on $M_\mathrm{Ni}$ and $E_\mathrm{exp}$ if they are discovered in future observations. Figure \ref{LCpeak} shows $t_\mathrm{max}$ and $L_\mathrm{max}$ for PISNe and AISNe. While $L_\mathrm{max}$ increases as a function of $M_\mathrm{He}$, $t_\mathrm{max}$ is not monotonous because $t_\mathrm{max}\propto(M_\mathrm{ej}^3/E_\mathrm{exp})^4$ and both $M_\mathrm{ej}$ and $E_\mathrm{exp}$ increase as a function of $M_\mathrm{He}$. It is seen from Fig.~\ref{LCpeak} that $t_\mathrm{max}$ for AISNe with $f=0.5$ is shorter than that for PISNe. This is because AISNe show larger $E_\mathrm{exp}$ than PISNe, as seen in Fig.~\ref{Eexp}.



\begin{figure}[t]
 \centering
 \includegraphics[width=8.5cm]
      {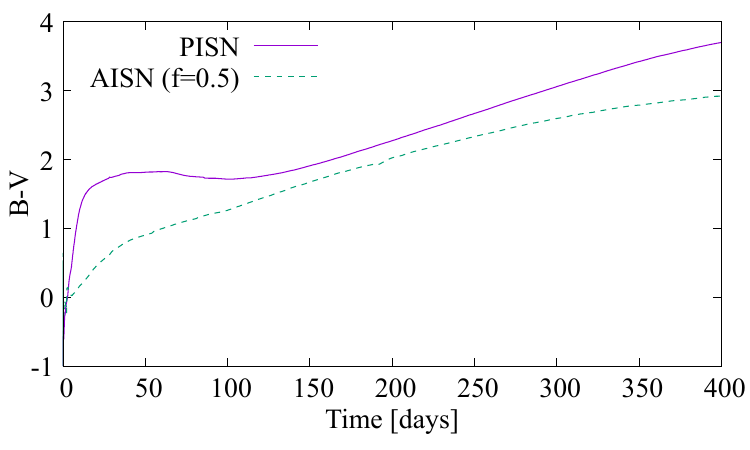}
 \caption{The color measured in $B-V$ of a PISN (upper panel) and an AISN (lower panel) with $f=0.5$.
The stellar mass is fixed to $M_\mathrm{He}=98M_\odot$.}
 \label{B-V}
\end{figure}

Since observational instruments adopt filters which transmit only a specific frequency band, the bolometric light curves cannot be directly compared to observations. In order to compare the models with future observations, we calculated multi-band light curves using a radiation hydrodynamics code STELLA \citep{1998ApJ...496..454B,2000ApJ...532.1132B,2006A&A...453..229B}. In Fig.~\ref{B-V}, we show the color measured in $B-V$. The plot shows that PISNe are redder than AISNe for a fixed stellar mass. At $t=t_\mathrm{max}$,  $B-V=1.81$ mag for the PISN and $1.42$  mag for the AISN. This is because the color is dependent on the ratio between $M_\mathrm{Ni}$ and $M_\mathrm{He}$ \citep{2012MNRAS.426L..76D}. In the two models shown in Fig.~\ref{B-V}, $M_\mathrm{He}$ is fixed but $M_\mathrm{Ni}$ is larger in the AISN model. When $M_\mathrm{Ni}$ is higher, heating induced by the $^{56}$Ni decay chain increases the photospheric temperature and leads to bluer light curves.

\subsection{Event Rates}

\begin{figure}[t]
 \centering
 \includegraphics[width=8.5cm]
      {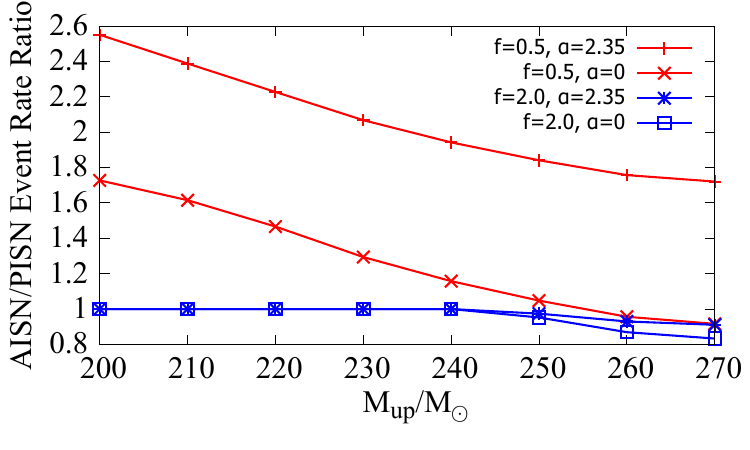}
 \caption{The event rate ratio $R$ as a function of the IMF cutoff $M_\mathrm{up}$. Two values of the IMF slope $\alpha=2.35$ and 0 are adopted.}
 \label{rate}
\end{figure}

Because the mass range for AISNe is lighter than that for PISNe, the event rates for these transients are different. Figure \ref{rate} shows the event rate ratio between AISNe and PISNe, estimated as
\begin{equation}
    R=\frac{\int^{\min(M_\mathrm{init, max}^\mathrm{ALP},\;M_\mathrm{up})}_{M_\mathrm{init, min}^\mathrm{ALP}}M^{-\alpha}dM}{\int^{\min(M_\mathrm{init, max},\;M_\mathrm{up})}_{M_\mathrm{init, min}}M^{-\alpha}dM},
\end{equation}
where $\alpha$ is the slope of the initial mass function (IMF), $M_\mathrm{up}$ is the high mass cutoff of the IMF,  and $M_\mathrm{init,\;max}$ $ (M_\mathrm{init,\;min})$ is the maximum (minimum) initial mass of PISNe and AISNe. The superscript ALP stands for quantities for AISNe. Although the typical value for $\alpha$ is 2.35 \citep{1955ApJ...121..161S}, hydrodynamical simulations for the formation of metal-free stars suggest top-heavy IMFs \citep[e.g.,][]{2014ApJ...792...32S,2014ApJ...781...60H,2015MNRAS.448..568H,2016MNRAS.462.1307S,2020ApJ...897...58T,2021MNRAS.508.4175C}. In Fig.~\ref{rate} we show two IMFs, a flat IMF with $\alpha=0$ to represent the top-heavy IMFs and the Salpeter IMF with $\alpha=2.35$.  \citet{2014ApJ...780..117S} analyzed very massive stars in the stellar cluster R136 and reported that $M_\mathrm{up}$ for metal-rich stars would be in the range $200$--$500M_\odot$. As for metal-free stars, stars heavier than $300M_\odot$ were not formed in   cosmological simulations performed by \citet{2014ApJ...792...32S}, while simulations in \citet{2014ApJ...781...60H} reported the formation of stars as heavy as $\sim1000M_\odot$. Since the maximum mass for the star formation is highly uncertain, $M_\mathrm{up}$ is adopted as a free parameter in Fig.~\ref{rate}.

Figure \ref{rate} shows that the PISN-like event rate is enhanced by a factor of $\sim1.7$--$2.6$ for $f=0.5$ and the Salpeter IMF. If the ALP mass is larger, the event rate enhancement is suppressed because the AISN mass range becomes closer to the PISN case. The event rate ratio is higher with smaller $M_\mathrm{up}$ because the PISN range extends to heavier masses compared with AISNe and the formation of such heavy stars is prevented by the IMF cutoff. Although it would be necessary to consider a realistic star formation history to predict the expected event number for surveys, the number of AISNe would be systematically higher than PISNe.

\subsection{Prospects for Constraining the ALP Mass with AISN Light Curves}
In Figs.~\ref{Ni} and \ref{Eexp}, we saw that the $^{56}$Ni mass and the explosion energy change as a function of $m_a$ for a fixed $M_\mathrm{He}$. It is thus expected that estimating $M_\mathrm{He}$, $M_\mathrm{Ni}$, and $E_\mathrm{exp}$ in future PISNe or AISNe would give a clue to distinguish the two scenarios. As Fig.~\ref{LCpeak} shows, they can be estimated from $t_\mathrm{max}$ and $L_\mathrm{max}$ of the light curve. Although light curves of PISN-like events have not been discovered, future observations are planned to find such events \citep{2018MNRAS.479.2202H,2020ApJ...894...94R,2022ApJ...925..211M,2022arXiv220408727M,2022arXiv220409402T}. Comparison between the AISN models and observed light curves would provide a constraint on $m_a$. For example, it is estimated that James Webb
Space Telescope can detect PISNe heavier than $200M_\odot$ (i.e. $M_\mathrm{He}\gtrsim98M_\odot$) out to redshift of $z\sim7.5$ \citep{2018MNRAS.479.2202H}. The light curve from a transient at $z$ is stretched by a factor of $1+z$ in the observer frame. Therefore, the difference $\Delta t_\mathrm{max}$ between the peak times of PISNe and AISNe  in the observer frame can be observationally distinguished if the observing cadence is shorter than  $(1+z)\Delta t_\mathrm{max}$. If we assume $z=7.5$, $(1+z)\Delta t_\mathrm{max}$ is between $\sim75$--150 days for the $f=0.5$ case, while it is $\sim0$--40 days for the $f=2.0$ case. It is hence possible to distinguish AISNe and PISNe  if  ALPs  are light enough. Also, the larger $M_\mathrm{Ni}$ for AISNe leads to the higher luminosity. If we adopt the PISN observational threshold $M_\mathrm{He}\gtrsim98M_\odot$ mentioned above, the analogous threshold for AISNe with $f=0.5$ would become as low as $M_\mathrm{He}\gtrsim77M_\odot$.

In our calculations, we assumed that the progenitor is hydrogen-free. If the helium core is surrounded by a thick envelope, the morphology of the light curves becomes more complex \cite[e.g.,][]{2011ApJ...734..102K}. In addition, the envelope makes it impossible to equate $M_\mathrm{ej}$ with $M_\mathrm{He}$. In order to extract the information of the hydrogenic progenitor, it is desirable to perform hydrodynamic modeling of the light curves. Also, it is notable that the nebular phase spectra of PISNe would provide information on $M_\mathrm{Ni}$ even for the hydrogenic progenitors \citep{2016MNRAS.455.3207J}.

It is known that the evolution of PISNe is sensitive to the $^{12}$C$(\alpha,\;\gamma)^{16}$O reaction rate \citep{2018ApJ...863..153T,2020ApJ...902L..36F}. In particular, when the rate is higher, the $^{56}$Ni production becomes more efficient. Because the low-energy cross sections of the reaction are still uncertain, it might be difficult to distinguish the effects of ALPs and the high reaction rate. Future experiments such as JUNA \citep{2022EPJWC.26008001L} may reduce the uncertainty in the reaction rate.

\section{Conclusion}

In this study, we explored the $^{56}$Ni mass and the event rates of axion instability supernovae (AISNe). Searches for PISN-like events including AISNe are being planned using next-generation near-infrared telescopes. Once a PISN/AISN candidate is discovered, its light curve and spectra would give us information on the $^{56}$Ni mass and the stellar mass. As we saw in Fig.~\ref{Ni}, these quantities can be used to distinguish PISNe and AISNe. Also, as discussed in \cite{2022PhRvD.105i5038S}, the mass distribution of astrophysical black holes would be affected by heavy ALPs. Optical observations of AISNe  and gravitational-wave observations of black hole binaries would hence provide complementary methods to probe the nature of ALPs.

An object of interest is the superluminous supernova PTF12dam \citep{2012ATel.4121....1Q}. It is argued that this object is not a PISN because of its short rise time \citep{2013Natur.502..346N}. Although this object has a shorter rise time than our AISN models, it would be worthwhile to perform sensitivity studies on the $^{12}$C($\alpha,\;\gamma)^{16}$O reaction rate because a larger rate would lead to a smaller $M_\mathrm{ej}$ for a fixed $M_\mathrm{Ni}$ and shorten the model rise time.

When a star is not massive enough to reach total disruption, it can cause the pulsational pair or the pulsational axion instabilities \citep{2022PhRvD.105i5038S}. It would be interesting to explore the models for these cases as well, because there is a candidate for a pulsational pair instability supernova \citep{2022arXiv220506386W} and more examples would be discovered by future observations.

Furthermore, nucleosynthesis in PISNe and AISNe may have left traces in elemental abundances of metal-poor stars. \cite{2014Sci...345..912A} reported that a very metal-poor star SDSS J001820.5–093939.2 shows an elemental composition that is similar to PISN yields.  Although such stars are extremely rare, it is worthwhile to perform post-process network calculations with a large nuclear reaction network to predict detailed nucleosynthesis yields of AISNe. 

In this calculation, we focused on photophilic ALPs because AISNe provide a unique way to probe an interesting parameter region called the cosmological triangle. However, new transients that are similar to PISNe can be induced by any new bosons if they are tightly coupled with the plasma \citep{2022PhRvD.105i5038S}. Extending the calculations to various particles would provide a general way to constrain physics beyond the Standard Model.

\begin{acknowledgments}
The authors thank Jeremy Sakstein and Djuna Croon for fruitful discussion. K.M.\ was supported by  Research Institute of Stellar Explosive Phenomena (REISEP) at Fukuoka University and JSPS KAKENHI Grant Number JP21K20369. K.K. was supported in part by Grants-in-Aid for Scientific Research of the Japan Society for the Promotion of Science (JSPS, No. JP22H01223), 
the Ministry of Education, Science and Culture of Japan (MEXT, Nos.
JP17H06364, JP17H06365), 
 by the REISEP (No. 207002),
and JICFuS as “Program for Promoting researches on the Supercomputer Fugaku” (Toward a unified view of 
the universe: from large scale structures to planets, JPMXP1020200109).
The work of S.H.\ is supported by the U.S.~Department of Energy Office of Science under award number DE-SC0020262, NSF Grant No.~AST1908960 and No.~PHY-1914409, and JSPS KAKENHI Grant Number JP22K03630. This work was supported by World Premier International Research Center Initiative (WPI Initiative), MEXT, Japan. S.B. is supported by RSCF grant 19-12-00229 in his work on SN
light curve simulations.
Numerical computations were  performed by the PC cluster at Center for Computational Astrophysics, National Astronomical Observatory of Japan.

\software{MESA \citep{Paxton2011,Paxton2013,Paxton2015,Paxton2018,Paxton2019}, Reproduction Package for the Paper ``Axion Instability Supernovae" \citep{2022PhRvD.105i5038S} and STELLA \citep{1998ApJ...496..454B,2000ApJ...532.1132B,2006A&A...453..229B}. The MESA inlists used in this study are available in Zenodo: \dataset[10.5281/zenodo.7288159]{https://doi.org/10.5281/zenodo.7288159}.}

\end{acknowledgments}


\bibliography{ref}{}
\bibliographystyle{aasjournal}



\end{document}